\documentclass[12pt,preprint]{aastex}
\shorttitle{Mg~II absorber in J0035+0114}
\shortauthors{Jiang et al.}

\begin{document}
\title{A Dusty Mg~II absorber Associated with the Quasar SDSS J003545.13+011441.2\footnote{Some of the data presented herein were obtained at the W.M. Keck Observatory, which is operated as a scientific partnership among the California Institute of Technology, the University of California and the National Aeronautics and Space Administration. The Observatory was made possible by the generous financial support of the W.M. Keck Foundation.}}
\author{P. Jiang\altaffilmark{1,2}, J. Ge\altaffilmark{2}, J. X. Prochaska\altaffilmark{3,4}, V. P. Kulkarni\altaffilmark{5}, H. L. Lu\altaffilmark{1}, and H. Y. Zhou\altaffilmark{1}}
\altaffiltext{1}{Key Laboratory for Research in Galaxies and Cosmology, The University of Science
and Technology of China, Chinese Academy of Sciences, Hefei, Anhui, 230026, China}
\altaffiltext{2}{Astronomy Department, University of Florida, 211 Bryant Space Science Center, P. O. Box 112055, Gainesville, FL 32611}
\altaffiltext{3}{University of California Observatories-Lick Observatory, University of California, Santa Cruz, CA 95064}
\altaffiltext{4}{Department of Astronomy and Astrophysics, University of California, Santa Cruz, CA 95064}
\altaffiltext{5}{Department of Physics and Astronomy, University of South Carolina, Columbia, SC 29208}
\email{jpaty@mail.ustc.edu.cn}

\begin{abstract}
We report on a dusty Mg~II absorber associated with the quasar SDSS
J003545.13+011441.2 (hereafter J0035+0114) at $z$=1.5501, which is
the strongest one among the three Mg~II absorbers along the sight
line of quasar. The two low redshift intervening absorbers are at
$z$=0.7436, 0.5436, respectively.
Based on the photometric and spectroscopic data of Sloan Digital Sky
Survey (hereafter SDSS), we infer the rest frame color excess E(\bv) due
to the associated dust is more than 0.07
by assuming a Small Magellanic Cloud (hereafter SMC) type extinction curve.
Our follow-up moderate resolution spectroscopic
observation at the 10-m Keck telescope with the ESI spectrometer
enable us to reliably identify
most of the important metal elements,
such as Zn, Fe, Mn, Mg, Al, Si,
Cr, and Ni in the associated system. We measure the column density of each species
and detect significant dust depletion.
In addition, we develop a simulation technique to gauge the significance of 2175-{\AA} dust absorption
 bump on the SDSS quasar spectra. By using it,
we analyze the SDSS spectrum of J0035+0114 for the presence of a associated 2175-{\AA} extinction
feature and report a tentative detection at $\sim$2$\sigma$ significant level.
\end{abstract}


\section{Introduction}
Absorption lines in the spectra of quasars have been detected
since shortly after the discovery of quasars (e.g. Burbidge,
Lynds \& Burbidge 1966; Schmidt 1966). They provide us with a powerful
tool to probe abundances, physical conditions and kinematics of gas
in a wide variety of environments. The absorption systems could be divided
into two populations by the difference between the redshift of absorption
lines ($z_{abs}$) and the redshift of quasar emission lines ($z_{em}$).
If $z_{abs}$ and $z_{em}$ are almost the same ($z_{abs}-z_{em} <$
5,000 km s$^{-1}$ in quasar rest frame), the absorption system is
the absorption system is taken to be associated with the quasar.
Otherwise the absorption system is usually intervening
(Weymann et al. 1979, Foltz et al. 1986), although some associated
C~IV absorbers may be found at relative velocities of as much as
75,000 km/s with respect to the quasar (Richards et al. 1999).

The presence of dust grains associated with varieties of absorbers could be
constrained by measuring the relative abundances of volatile and
refractory elements in order to infer a dust depletion level. The depletion
of Cr with respect to Zn in intervening Damped Ly$\alpha$ Absorption systems
(DLAs) shows the existence of
dust in the high density gas (e.g. Pettini et al. 1994, 1997, 1999; Nestor et al. 2003)
All of the DLAs showing high dust depletion level, [Zn/Fe]$>$0.8, have
high molecular hydrogen detection rate (e.g.
Ge \& Bechtold 1997; Ge et al. 2001; Cui et al. 2005; Noterdaeme et al. 2008).

Dust in an absorber could also be inferred by measuring
reddening and extinction effects on the background objects.
York et al. (2006) studied the extinction of 809 intervening Mg~II
absorption systems from the SDSS in a
statistical way at $1\le z_{abs}<2$.
Their extracted average extinction curves were similar to the
SMC curve with E(\bv)$\le0.08$, and indicated a tentative correlation
between E(\bv) and the Mg~II equivalent width in the rest
frame $W_0^{\lambda2796}$.
Recently, M\'{e}nard et al. (2008) composed
a much larger sample with almost 7000 strong Mg~II absorption systems
at $0.4 < z_{abs}<2.2$ and confirmed the correlation between
E(\bv) and $W_0^{\lambda2796}$.

The prominent difference between extinction curves of Milky Way (MW)
and that of the SMC is the presence of a 2175-{\AA} dust absorption bump
(Savage \& Mathis 1979; Fitzpatrick 1989). Reliable detections of the
2175-{\AA} feature in individual intervening absorption system are
rare. Cohen et al. (1999) detected the 2175-{\AA} feature in a
damped Ly$\alpha$ absorber at a redshift of z=0.524 toward the
BL Lac object AO 0235+164 (later updated by Junkkarinen et al. 2004). And
Wang et al. (2004) identified three intervening Mg II absorption systems
at $1.4 < z < 1.5$ with the 2175-{\AA} dust absorption
feature in quasar spectra from the SDSS.
Srianand et al. (2008) found a 2175-{\AA} extinction feature in two Mg~II
systems at z$\sim$1.3 and detected 21-cm absorption in both of them, which
usually traces cold dense gas content. Recently, Noterdaeme et al (2009)
presented a detection of carbon monoxide molecules (CO) at $z$ = 1.6408
towards a red quasar and a pronounced 2175-{\AA} bump at the redshift
of CO absorber. In the past several years, analysis of GRB afterglow spectra has also 
revealed several positive detections from intervening absorbers and from
gas in the GRB host galaxies (e.g. Ellison et al. 2006; El{\'i}asd{\'o}ttir et al. 2009;
Prochaska et al. 2009).

In this paper, we report on a dusty Mg~II absorber associated with the quasar
SDSS J0035+0114 at $z$=1.5501. In the next section, we
will describe the observation data, including SDSS spectrum and
our follow-up spectroscopic
observation at the 10-m Keck telescope with ESI spectrometer.
In \S3, we infer the color excess E(\bv) of dust extinction with
SDSS photometric and spectroscopic data.
In \S4, we measure the column densities of
important metal ions with high accuracy on Keck spectrum and explore the dust
depletion patterns in the absorber.
The possible detection of 2175-{\AA} absorption bump will
be disscused in \S5. Our main results will be
summarized in the last section, together with a discussion.

\section{Observations}
The SDSS images of J0035+0114 were acquired on UT 2001 Oct 15. The
point-spread function magnitudes measured from the
images are $19.997\pm 0.040$, $19.293\pm 0.010$, $18.972\pm 0.010$,
$18.493\pm 0.012$, and $18.344\pm 0.048$ in $u$, $g$, $r$, $i$, and $z$,
respectively. The SDSS spectrum was obtained on UT 2000 Sep 06 and
covers $\sim 3800-9200$ \AA~with a spectral resolution $R \sim
2000$ and a median S/N~$\approx 7$ (Stoughton et al. 2002).
The SDSS spectrum is remarkable with relatively red color and strong
associated absorption lines imposed on it.
Because our initial inspection suggested a very dusty system,
we performed follow-up spectroscopic observations of the quasar at higher
spectral resolution.
On UT 2004 Sep 11, we acquired two 1200s exposures
of J0035+0114 with the ESI spectrometer (Sheinis et al. 2002)
on the 10m Keck~II telescope. We employed the 0.5$''$ slit providing a
FWHM$\approx 37$ km s$^{-1}$
resolution and a wavelength coverage $\lambda = 4000 - 10,000$\AA. The
spectral images were reduced and calibrated using the
XIDL\footnote{http://www.ucolick.org/$\sim$xavier/IDL} software package
ESIRedux\footnote{http://www2.keck.hawaii.edu/inst/esi/ESIRedux/index.html}
(v1.0).  The optimally extracted 1D
spectra were converted to vacuum wavelengths and converted to the
heliocentric frame and then flux-calibrated using a
spectrophotometric standard acquired that same night.  The data
were normalized by fitting a series of polynomials to absorption-free  
regions of the quasar spectrum. The emission redshift of $z=1.5501$ is
measured.

\section{Color and Reddening}
The color of quasars is redshift dependent, since the broad emission
features on underlying continuum move in/out the photometric
passbands at different redshift (Richards et al. 2001). Richards et al. (2003) introduced
a relative color to determine the underlying continuum color of quasars
by subtracting the median colors of quasars at the redshift of each
quasar from the measured colors of each quasar. The relative color
$\Delta (g-i)$ can be used to distinguish between reddened quasars
and optically steep quasars.
The distribution of relative colors should be a Gaussian, assuming
a Gaussian distribution of power-law spectral indices of quasars.
However, $\Delta (g-i)$ shows a significant
asymmetric tail to the red end. The objects in this tail are reddened by
dust.

The relative color $\Delta (g-i)$ of J0035+0114 is 0.40$\pm$0.02.
$\Delta (g-i)$s of all quasars with redshift between 1.525 and 1.575
in SDSS Data Release 7 (Abazajian et al. 2009)
are extracted to compose the relative color distribution 
at the redshift of J0035+0114 ($z$=1.5501). All the colors in this analysis
are dereddened by using the dust map of Schlegel et al. (1998).
In Figure 1, it is clear that J0035+0114 is in the red tail of the composed
color distribution.
By assuming the intrinsic spectral index of J0035+0114 is
flat ($\Delta (g-i)$=0), we infer that it could be reddened by
an SMC type extinction curve with E(\bv)$\sim0.09$ in the rest frame of
quasar emissions at $z$=1.5501 (Richards et al. 2003).

On the Keck spectrum of J0035+0114, 
three absorption line systems can be readily identified
, with redshifts of z=1.5501, 0.7436, and
0.5436, respectively (hereafter system-A, -B, and -C).
The Mg~II$\lambda \lambda$2796,2803 doublet
has been detected in all of the three systems.
The Mg~II$\lambda$2796 absorption line in system-A is the strongest
with rest equivalent width $W_0^{\lambda 2796}=1.986\pm0.022$
($W_0^{\lambda 2796}=1.622\pm0.186$ in system-B;
$W_0^{\lambda 2796}=0.923\pm0.182$ in system-C).
To examine the dust reddening on SDSS spectrum of J0035+0114, we fit it\footnote{
The SDSS spectrum has been corrected for the Galactic reddening of E(\bv)=0.022 before
fitting procedure.} with two reddened composite SDSS quasar spectrum models.
First, we assume the dust reddening is solely due to system-A, which is the strongest
one. The composite SDSS spectrum (Vanden Berk et al. 2001)
is reddened by SMC type extinction curve (Pei 1992), in which E(\bv)
is a free parameter and $R_{V}=2.93$ is fixed, at $z$=1.5501.
To focus on fitting continuum of quasar spectrum, the regions with strong emission lines
and known strong absorption lines are masked.
The best fitted E(\bv) is 0.15, with $\chi^{2}_{\nu}$=1.28 (see Figure 2a).
Second, we fit the SDSS spectrum with a three absorbers model. We assume the total
dust reddening is contributed by three SMC extinction curves with the same E(\bv)
at the redshifts of 1.5501, 0.7436, 0.5436. The best fitted E(\bv) is 0.07, with
$\chi^{2}_{\nu}$=1.28 (see Figure 2b). Since system-A is the strongest absorber,
we infer that its rest frame E(\bv) would be greater than 0.07, which is the average
reddening of the three absorbers, by assuming E(\bv) scales with
$W_0^{\lambda 2796}$ in strong quasar Mg~II absorption systems (M\'{e}nard et al. 2008).

\section{Column Density and Dust Depletion}
Most of the important heavy elements are reliably identified in system -A
(see Figure 3).
Column densities of all elemental ions except Zn$^+$
were first estimated by measuring the apparent optical depth (AOD,
Savage \& Sembach 1991). Zn~II$\lambda \lambda$2026,2062
are heavily blended with Mg~I$\lambda$2026 and Cr~II$\lambda$2062.
To measure the column density of zinc, we first fitted
Mg~I$\lambda$2852 with a single Voigt profile, and then fitted
Mg~I$\lambda$2026, Zn~II$\lambda \lambda$2026,2062, and Cr $\lambda
\lambda$2056,2062,2066 simultaneously using the modeled profile of
Mg~I$\lambda$2852 as a template for each line. The multi-Voigt
fitting gave a column density of log N(Zn~II)=$13.21 \pm 0.03$. 
In addition, we fitted the Fe~II multiplets
with Voigt profile fitting of multiple lines. The fitting of Fe~II
$\lambda\lambda$2249,2260 together gave a column density of log
N(Fe~II)=$15.2\pm0.04$. While the
fitting of Fe~II $\lambda \lambda$2344,2374,2382 together gave a value of
$14.88\pm0.03$.
The equivalent width and column density
measured from ionized lines are presented in Table 1.
We also measure the column densities of strong Mg~II and Fe~II absorption lines
in system-B and system-C (see Table 2 and Figure 4). 

``The expanded SDSS/HST sample of low-redshift DLAs", compiled by
Rao et al. (2006), shows the success rate of DLA detection is
42\% $\pm$ 7\% for strong Mg~II--Fe~II systems with
$W_0^{\lambda 2796} / W_0^{\lambda 2600} < 2$ and Mg~I
$W_0^{\lambda 2852} \ga 0.1$ {\AA}. The associated Mg~II absorber here is right
in this high detection rate region and it may therefore be a dusty DLA absorption
system. However, it is at low redshift $z<1.65$,
where the Ly$\alpha$ transition cannot be observed in optical band
with ground instruments. Thus, the column density of neutral hydrogen
$N$(H~I) cannot be measured.
Since there are no lines of high ions detected
in the Keck spectrum (except for Al~III), we assume that hydrogen gas is
mostly neutral and other heavy elements are singly ionized in this Mg~II
absorption system while measuring column densities and dust depletion levels.

The presence of dust grains could be
constrained by measuring the relative abundances of volatile and
refractory elements in order to infer a depletion level of metal in gas phase.
As zinc is a relatively undepleted element due to its low
condensation temperature, the ratios of other heavy elements
to it are usually used to measure the dust depletion (Meyer \& Roth 1990; Pettini et al. 1994).
The depletion factors, [X/Zn]=log(N(X)/N(Zn))-log(N(X)/N(Zn))$_{\sun}$, are listed in Table 3.
It is clear that the absorber associated with J0035+0114 has significant
depletion factors\footnote{Zn~II$\lambda \lambda$2026,2062 of system-B and
system-C are not covered by the Keck spectrum. Thus, the dust depletion patterns in
those systems are not obtainable.},
which indicates that this system contains substantial dust grains.
Figure 5 plots the dust depletion patterns compared with
the ``warm" and ``cold" Galactic interstellar medium
(ISM) and SMC ISM. It seems that the depletion pattern in this absorber is
similar to those found in sight lines through ``warm" gas
of MW. 

\section{Possible 2175-{\AA} Absorption Bump}
Initially the quasar absorption line system in J0035+0114 at $z$=1.5501
is a candidate in our ongoing project to search for quasar absorption
line systems with 2175-{\AA} absorption bump feature (Zhou et al. 2010).
It was selected because the suppressed flux of quasar around 2200-{\AA}
in the rest frame of the associated Mg~II absorber. However, the
depression can be caused by the variation of the strength of Fe~II emissions
, too (Pitman et al. 2000). Hence, we develop a simulation technique to
gauge the significance of 2175-{\AA} dust absorption
bump on the SDSS quasar spectra.

J0035+0114 is taken as an example to introduce our simulation procedures below.
First, all the SDSS spectra of quasar with emission redshift in the range of
$z_{J0035}$-0.05 and $z_{J0035}$+0.05 with median S/N$>6$ are chosen
to compose a control sample and then are corrected for Galactic reddening.
We basically fit each of them by reddening the composite quasar spectrum
(Vanden Berk et al. 2001) with a parameterized extinction curve at redshift
of the absorber of interest.
The extinction curve is defined in a similar formula with the prescription of
Fitzpatrick \& Massa (1990), as
\begin{equation} 
A(\lambda)=c_1+c_2x+c_3D(x,x_0,\gamma)
\end{equation}
where $x=\lambda^{-1}$.
And $D(x,x_0,\gamma)$ is a Lorentzian profile, which is expressed as
\begin{equation}
D(x,x_0,\gamma)={\frac{x^2}{(x^2-x_0^2)^2+x^2\gamma^2}}
\end{equation}
where $x_0$ and $\gamma$ is the peak position and FWHM
of Lorentzian profile, respectively. Our aim is to unveil the 2175-{\AA}
absorption feature associated with absorption line systems on quasar spectra.
We do not try to derive the absolute extinction curve and our derived one
is a relative extinction curve without normalization. We cannot measure the
conventional extinction parameters $A_V$, E(\bv) and $R_V$ from it.
But all the features of 2175-{\AA} absorption bump are preserved.
The linear component in the extinction
curve accounts for the variation of quasar spectral index. Thus, the parameter
$c_2$ could be negative if a quasar spectrum is steeper than composite spectrum.
The Lorentzian profile is used to model absorption bump. The strength of bump is
measured by the area of bump $A_{bump}=\pi c_3/(2\gamma)$.\footnote{The area of bump defined
in this work is different from that in Fitzpatrick \& Massa (2007).
$A_{bump}$=E(\bv)$\times A_{bump}^{*}$, where $A_{bump}^{*}$ is the area defined in FM07.
$A_{bump}$ can be interpreted
as rescaling the integrated apparent optical depth of bump absorption
($A_{\lambda}={\frac{2.5}{ln 10}}\tau_{\lambda}$).}
However, this strength is not necessarily zero
even if the spectrum does not have any absorption bump feature. 
In the absence of a bump, the distribution of the best fitted strength is expected to be Gaussian
by assuming random fluctuation of Fe~II emission on each spectrum and photon noise.
During the fitting procedure, $x_0$ is fixed to 4.59 $\mu m^{-1}$ in the rest frame
of the absorber of interest and the width $\gamma$ is fixed
to 0.89 $\mu m^{-1}$ in the same frame.
\footnote{The most likely values of peak and width for
Galaxy 2175-{\AA} absorption bump are 4.59 $\mu m^{-1}$ and 0.89 $\mu m{^-1}$
(Fitzpatrick and Massa, 2007).} The three free parameters are $c_1$, $c_2$ and $c_3$.
To focus on fitting continuum of quasar spectrum, the regions with strong emission lines
(Mg~II$\lambda$2800, C~III]$\lambda$1909, C~IV$\lambda$1550, Si~IV$\lambda$1400) and
known strong absorption lines are masked.

The SDSS spectrum of J0035+0114 is modeled by a composite quasar spectrum
reddened using the parameterized extinction curve at the same redshift of it.
Although J0035+0114 could be reddened by three dusty absorbers simultaneously,
the two low redshift absorbers (system-B and -C) cannot contribute to the possible
2175-{\AA} extinction bump. In addition, we think the reddening effects of them can be
well modeled by the linear component of the parameterized extinction curve.
The fitting results are presented in Figure 6:
panel (a) shows the best model compared with observation data;
panel (b) shows the best fitting extinction curve in the quasar rest frame.
The goodness of fit is measured by $\chi^2_{\nu}=1.21$.
Then, 4576 quasar SDSS spectra around the redshift of J0035+0114 ($z$=1.5501) are selected
to compose a fairly big control sample. The histogram of fitted strength of bumps
is plotted in Figure 7. It can be fitted by using a single Gaussian function with 
the standard deviation $\sigma=0.08$. The strength of bumps extracted on the
spectrum of J0035+0114 is 0.15$\pm$0.02. In sense of statistics, the possibility
that this bump is a real absorption feature is $\sim$95\% (significant at $\sim$2$\sigma$ level).
Therefore, the 2175-{\AA} absorption bump associated with J0035+0114 is only a
marginal detection. In contrast, we measure six significant 2175-{\AA} extinction bumps
detected on SDSS spectra in literatures with the same formalism and find that they are
at least five times stronger than the one in this work (see Table 4).
The wavelength coverage of
SDSS spectrum is from 3800 {\AA} to 9200 {\AA}. It allows us to detect MW-like 2175-{\AA}
extinction feature up to redshift $z \sim 3$.
The highest redshift for identified 2175-{\AA}
bump to date is $z = 3.03$, which was detected by Prochaska et al. (2009) using
the afterglow Optical/IR photometry of gamma-ray burst GRB 080607.
Another high redshift 2175-{\AA} bump was
detected by El{\'i}asd{\'o}ttir et al. (2009) on the
afterglow spectrum of GRB 070802 at $z = 2.45$.

\section{Discussion and Summary}
The strongest Mg~II absorption system presented in this work is probably associated with
quasar, since it has ($z_{em}-z_{abs}) \sim$30 km s$^{-1}$ in quasar rest frame.
It is very possible that the
corresponding cold gas rises from the host galaxy of quasar.
However, we cannot rule out a nearby foreground galaxy absorption scenario.

Interstellar dust grains play an important role in the evolution of
galaxies, star formation and planet formation. 
The dust content in physical environment of quasar has not been well studied
(Li 2007 and reference therein).
More detection of associated dusty absorption system will aid us to understand
the nature of this dust and provide more clues on the evolution of galaxies and
quasars. Furthermore, the
population of dust-reddened quasars is a possible contributor to the X-ray background
(e.g. Shanks et al. 1991; Mushotzky et al. 2000; Brandt et al. 2000; Dong et al. 2005).
Its contribution is independent of the
contribution of type 2 quasars with completely obscured broad emission line
regions (Antonucci 1993).

On the basis of Keck spectroscopic observations,
the dust depletion factor,
[Fe/Zn]$\sim$-0.86, represents the high depletion population 
in high-redshift DLAs reported in Noterdaeme et al. (2008).
All of the previous DLAs showing high dust depletion factor,
[Fe/Zn]$<$-0.8, have high molecular hydrogen abundance.
Thus, this system is very likely
associated with high molecular hydrogen content. Unfortunately,
the molecular hydrogen absorption bands associated with this
system are in the UV region beyond the atmospheric transmission window.

Figure 8 plots the histogram of strength of 2175-{\AA} absorption bump measured on
328 Galactic extinction curves (Fitzpatrick \& Massa 2007). The dash lines on left
is the 3$\sigma$ threshold and the one on right is the 5$\sigma$ threshold suggested
by our simulation. If we only accept the detection at confidence level $\ge 5\sigma$,
more than $94\%$ of Galactic bumps can be recovered with our approach.
The success rate is so promising that we can develop a sensitive method for detecting
MW-like 2175-{\AA} extinction bumps on quasar spectra with Mg~II absorption lines in SDSS
database (Jiang et al. in preparation). The strength
of bumps measured in SMC (Gordon et al. 2003) are also plotted in Figure 8. The extinction
curve measured in the sight line of SMC wing exhibits a significant 2175-{\AA} absorption
bump. 

In summary, we identify three Mg~II absorption line systems along the sight line of
the quasar J0035+0114. The strongest one is most likely associated with the quasar. 
The dust content in this associated system is firmly detected
by either reddening or elements dust depletion pattern. The extinction curve of this system
is more likely to be SMC type with E(\bv) $>$ 0.07.
However, we detect a tentative 2175-{\AA} extinction bump at
$\sim$2$\sigma$ significant level with our parameterized extinction curve technique.
The high dust content suggest
this system is likely a DLA with molecular hydrogen content.

\acknowledgments
This work was partially supported by NSF with grant
NSF AST-0451407, AST-0451408 \& AST-0705139
and the University of Florida.
PJ acknowledges support from China Scholarship Council.
This research has also been partially supported by the
CAS/SAFEA International Partnership Program for Creative Research Teams.
VPK acknowledges support from NSF grant
AST-0607739 to the University of South Carolina.

The authors wish to recognize and acknowledge the very significant cultural
role and reverence that the summit of Mauna Kea has always had within the
indigenous Hawaiian community. We are most fortunate to have the opportunity
to conduct observations from this mountain.

Funding for the SDSS and SDSS-II has been provided by
the Alfred P. Sloan Foundation, the Participatings
Institutions, the National Science Foundation,
the U.S. Department of Energy, the National Aeronautics
and Space Administration, the Japanese Monbukagakusho,
the Max Planck Society, and the Higher Education
Funding Council for England.
The SDSS Web Site is http://www.sdss.org/.

The SDSS is managed by the Astrophysical Research
Consortium for the Participating Institutions. The
Participating Institutions are the American Museum
of Natural History, Astrophysical Institute Potsdam,
University of Basel, University of Cambridge,
Case Western Reserve University, University of Chicago,
Drexel University, Fermilab, the Institute for Advanced
Study, the Japan Participation Group, Johns Hopkins
University, the Joint Institute for Nuclear Astrophysics,
the Kavli Institute for Particle Astrophysics and
Cosmology, the Korean Scientist Group, the Chinese
Academy of Sciences (LAMOST), Los Alamos National
Laboratory, the Max-Planck-Institute for Astronomy
(MPIA), the Max-Planck-Institute for Astrophysics
(MPA), New Mexico State University, Ohio State University,
University of Pittsburgh, University of Portsmouth,
Princeton University, the United States Naval Observatory,
and the University of Washington.

\clearpage
\begin{figure}
\epsscale{1.0}
\plotone{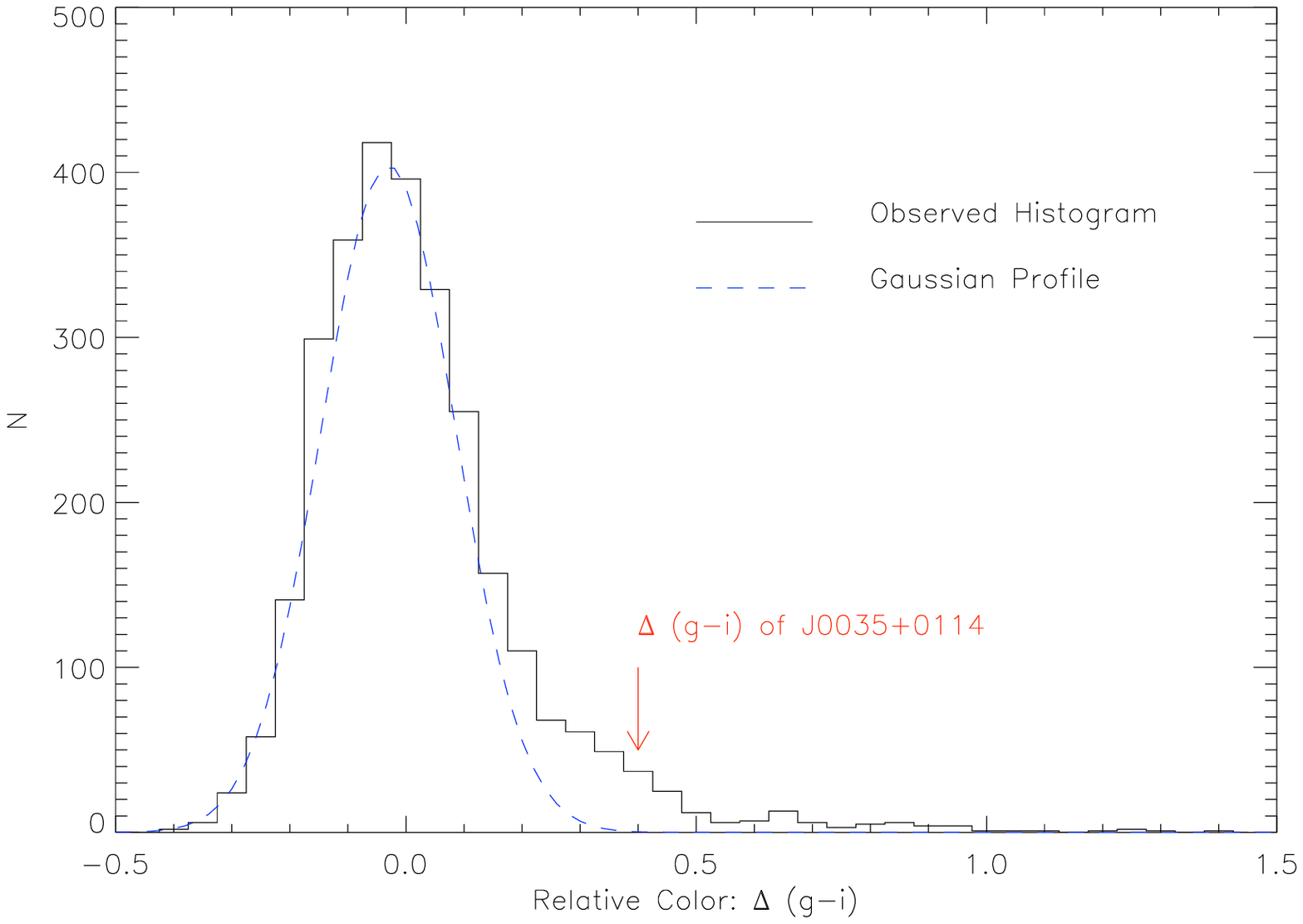}
\caption{The histogram of the relative color $\Delta(g-i)$ of SDSS quasar at redshift range between 1.525 and 1.575. Dashed blue line is a Gaussian fitting the blue wing. The red arrow indicates the relative color of J0035+0114.
\label{fig1}}
\end{figure}
\clearpage

\begin{figure}
\epsscale{1.0}
\plotone{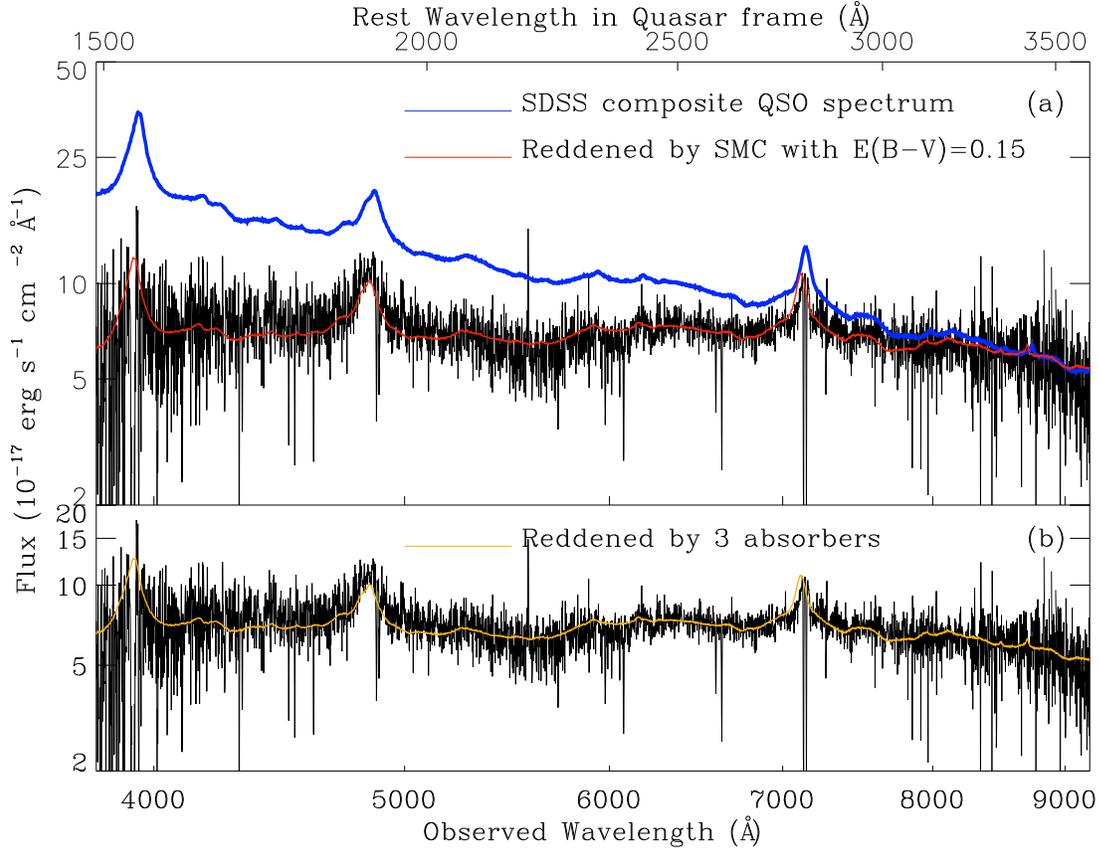}
\caption{In panel (a), the reddened composite quasar spectrum by using SMC extinction curve with E(\bv)=0.15 at z=1.5501 (the red line) overplotted with SDSS spectrum and the rescaled SDSS composite quasar spectrum (the blue line). The composite spectrum is scaled to match the observed data around 9000 \AA. In panel (b), the three absorbers model, composite spectrum is reddened by three SMC extinction curves with E(\bv)=0.07 at z=1.5501, 0.7436, 0.5436 (the orange line).
\label{fig2}}
\end{figure}

\begin{figure}
\epsscale{0.7}
\plotone{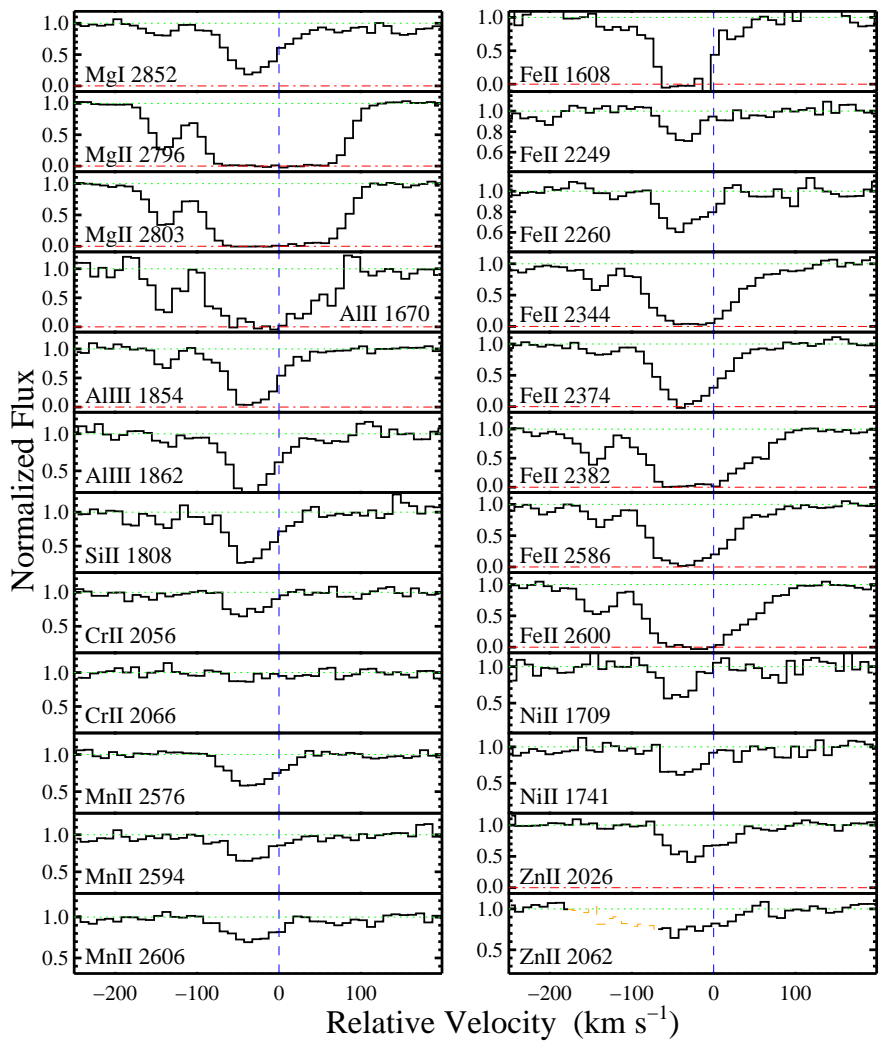}
\caption{Normalized Spectrum of J0035+0114 taken at the 10-m Keck telescope with ESI
spectrometer.
Strong absorption lines of ionized Fe, Ni, Si, Al, Cr, Mn, Mg, Zn in the
associated absorber are plotted in
velocity space. The dash lines indicate the $v=0$ km s$^{-1}$ at redshift 1.5501.
\label{fig3}}
\end{figure}

\begin{figure}
\epsscale{0.7}
\plotone{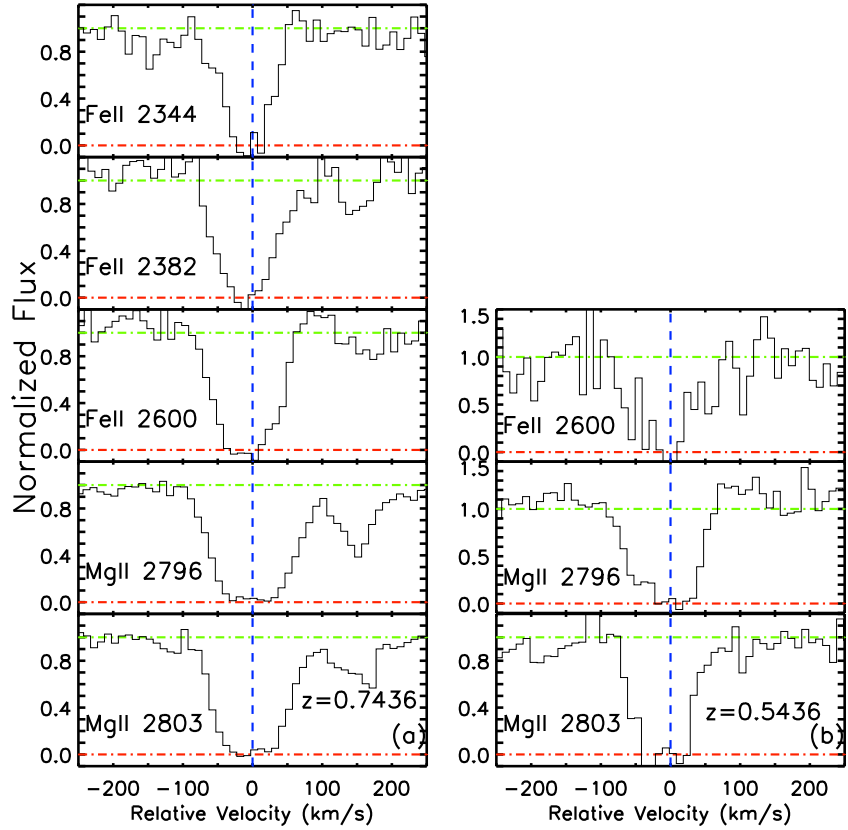}
\caption{Normalized Spectrum of J0035+0114 taken at the 10-m Keck telescope with ESI
spectrometer. Panel (a) is the strong absorption lines detected in the intervening absorber at z=0.7436; Panel (b) is the strong absorption lines detected in the intervening absorber at z=0.5436.
The dash lines indicate the $v=0$ km s$^{-1}$ in the rest frame of interested absorber.
\label{fig4}}
\end{figure}

\begin{figure}
\epsscale{1.}
\plotone{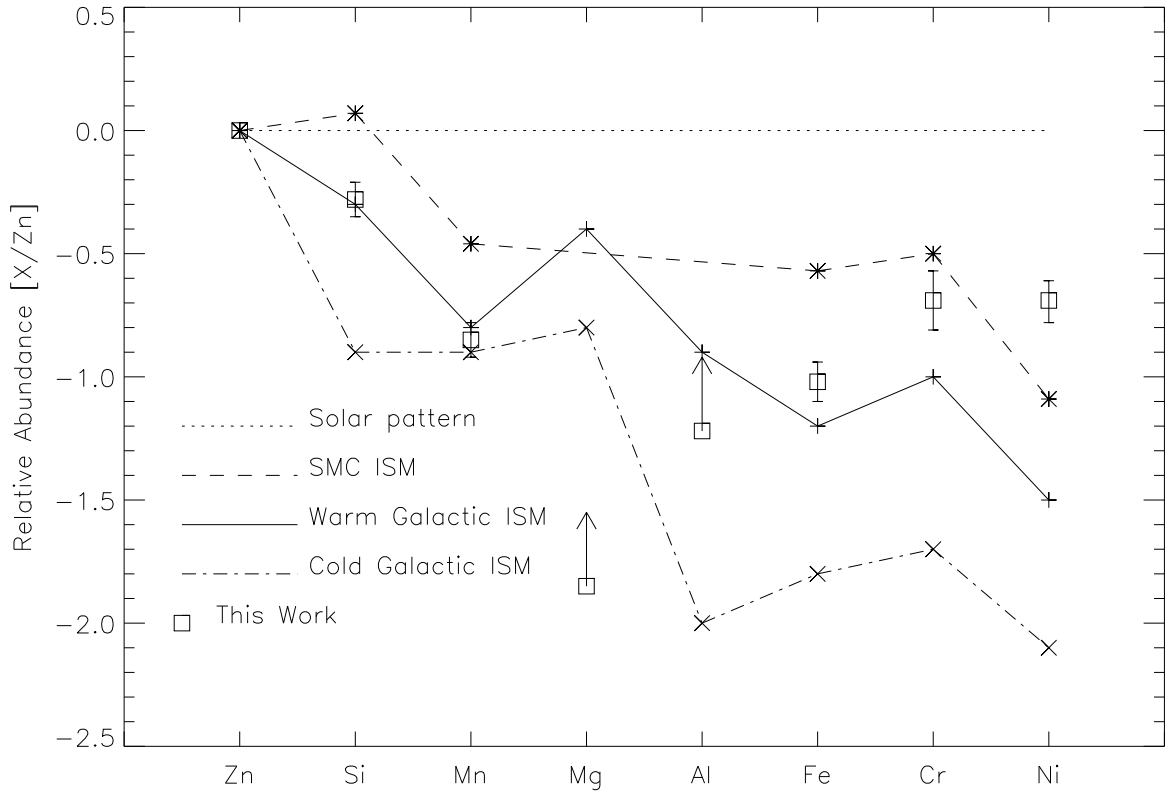}
\caption{Relative abundances are measured in J0035+0114 and compared with other known dusty clouds.
Values for ``warm" and``cold" Galactic ISM and SMC ISM were adopted form Jenkins et al. 1986, Welty et al. 1999 and Welty et al. 2001.
\label{fig5}}
\end{figure}

\begin{figure}
\epsscale{1.}
\plotone{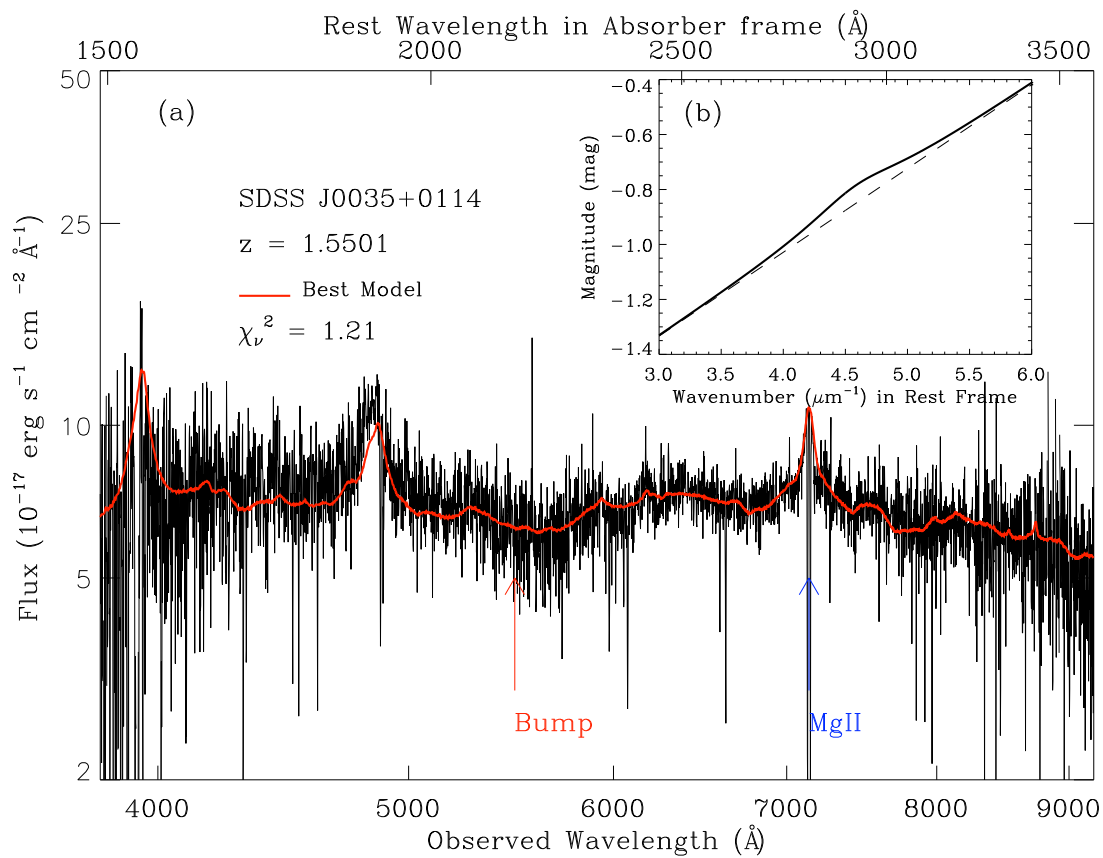}
\caption{In panel (a), the best fitted model is plotted with the observed data in the frame of observer. The red arrow indicates the center of fitted 2175-{\AA} absorption bump and the blue arrow indicates the Mg~II absorption lines.
In panel (b), the solid line is the best fitted extinction curve in rest frame of absorber. Its linear component is plotted with dashed line.
\label{fig6}}
\end{figure}

\begin{figure}
\epsscale{1.}
\plotone{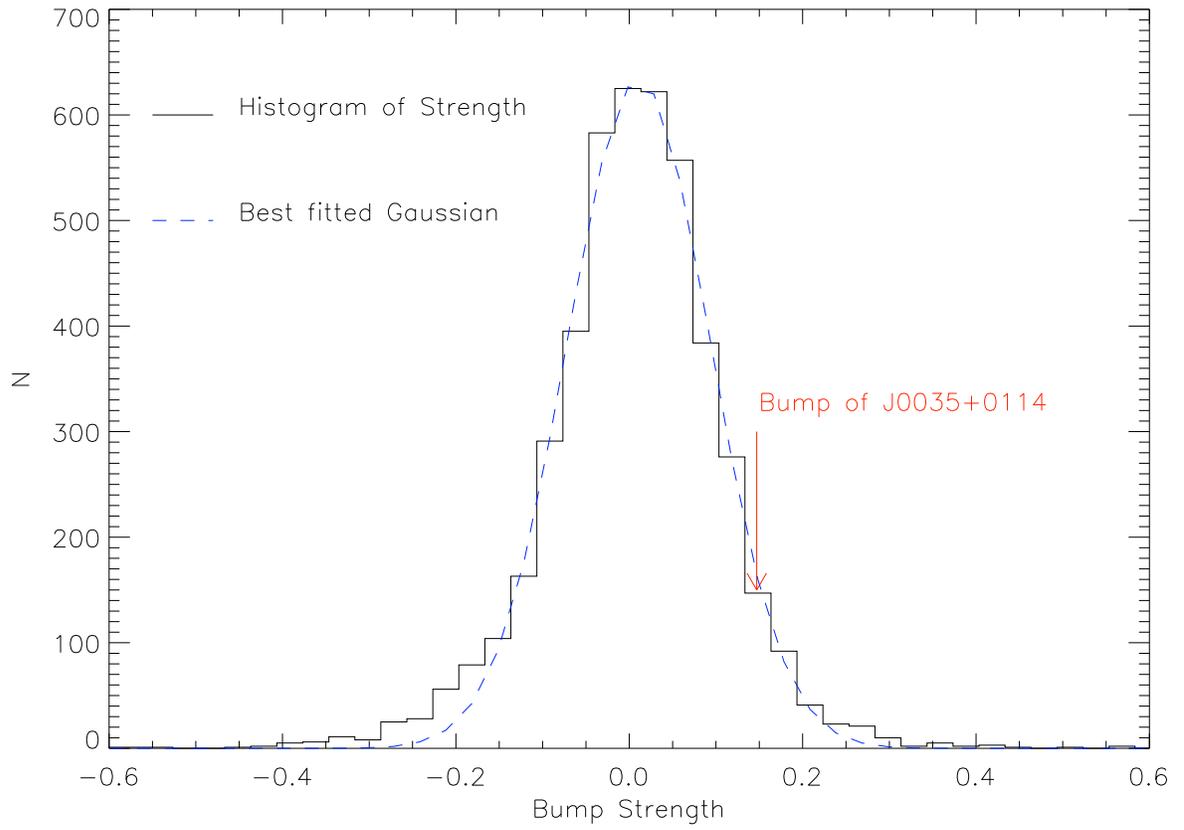}
\caption{The histogram of strength of bumps extracted in the simulation is presented in black line. The dashed blue line is the best fitted Gaussian. The red arrow indicates the strength of bump in J0035+0114.
\label{fig7}}
\end{figure}

\begin{figure}
\epsscale{1.}
\plotone{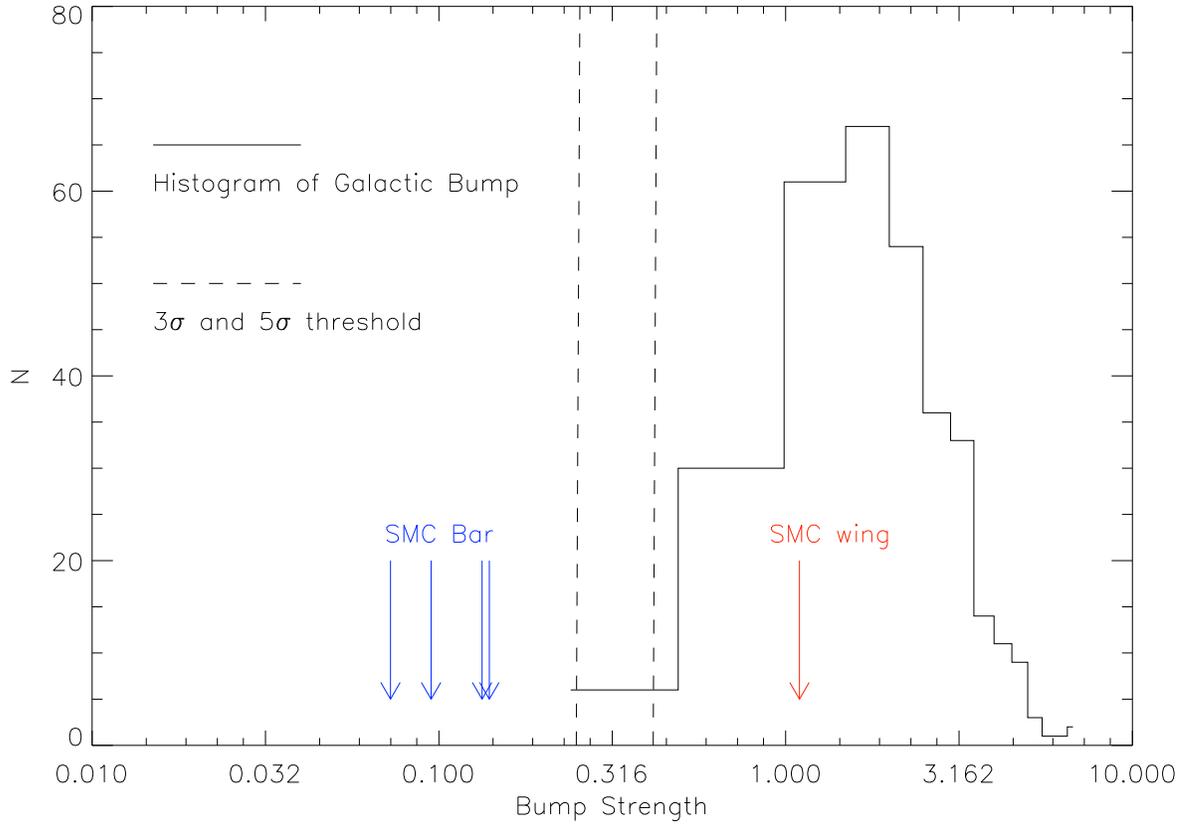}
\caption{The histogram of strength of 2175-{\AA} absorption bumps measured on
328 Galactic extinction curves (Fitzpatrick \& Massa 2007). The dash lines are 
the 3$\sigma$ and 5$\sigma$ thresholds suggested by the simulation. Blue arrows
indicate the bump strength measured in sight lines of SMC bar and the red arrow
indicated the bump strength measured in sight line of SMC wing (Gordon et al 2003).
\label{fig8}}
\end{figure}

\clearpage

\begin{deluxetable}{llccccc}
\tabletypesize{\scriptsize}
\tablecaption{Associated Strong Absorption Lines and Column Density measured by Apparent Optical Depth\label{tbl-1}}
\tablewidth{0pt}
\tablehead{
\colhead{$\lambda_{vacuum}$} & \colhead{Ion} & \colhead{$f$} & \colhead{$\lambda^a$} & \colhead{EW$^a$} & \colhead{N$_X^b$} & \colhead{N$_X^c$}\\
\colhead{(\AA)} & \colhead{} & \colhead{} & \colhead{(\AA)} & \colhead{(\AA)} & \colhead{log(cm$^{-2}$)}& \colhead{log(cm$^{-2}$)}
}
\startdata
1608.4511 & Fe \scriptsize{\uppercase\expandafter{\romannumeral2}} & 0.0580 & 1608.437 & 0.516$\pm$0.043 & 14.98$^{+0.02}_{-0.02}$ & ...\\
1670.7874 & Al \scriptsize{\uppercase\expandafter{\romannumeral2}} & 1.8800 & 1670.822 & 0.883$\pm$0.046 & 13.71$^{+0.02}_{-0.01}$ & ...\\
1709.6042 & Ni \scriptsize{\uppercase\expandafter{\romannumeral2}} & 0.0324 & 1709.558 & 0.103$\pm$0.022 & 14.18$^{+0.06}_{-0.07}$ & ...\\
1741.5531 & Ni \scriptsize{\uppercase\expandafter{\romannumeral2}} & 0.0427 & 1741.591 & 0.124$\pm$0.022 & 14.11$^{+0.05}_{-0.07}$ & ...\\
1808.0130 & Si \scriptsize{\uppercase\expandafter{\romannumeral2}} & 0.0022 & 1808.035 & 0.278$\pm$0.026 & 15.84$^{+0.03}_{-0.03}$ & ...\\
1854.7164 & Al \scriptsize{\uppercase\expandafter{\romannumeral3}} & 0.5390 & 1854.704 & 0.520$\pm$0.035 & 13.80$^{+0.02}_{-0.01}$ & ...\\
1862.7895 & Al \scriptsize{\uppercase\expandafter{\romannumeral3}} & 0.2680 & 1862.805 & 0.401$\pm$0.036 & 13.86$^{+0.03}_{-0.03}$ & ...\\
2026.136 & Zn \scriptsize{\uppercase\expandafter{\romannumeral2}} & 0.4890 & 2026.277$^d$ & 0.286$\pm$0.022 & 13.30$^{+0.03}_{-0.03}$ & 13.21$^{+0.03}_{-0.03}$\\
2026.4768 & Mg \scriptsize{\uppercase\expandafter{\romannumeral1}} & 0.1120 & 2026.277$^e$ & 0.286$\pm$0.022 & 13.95$^{+0.03}_{-0.03}$ & ...\\
2056.2539 & Cr \scriptsize{\uppercase\expandafter{\romannumeral2}} & 0.1050 & 2056.236 & 0.123$\pm$0.017 & 13.57$^{+0.04}_{-0.05}$ & ...\\
2062.234 & Cr \scriptsize{\uppercase\expandafter{\romannumeral2}} & 0.0780 &2062.563$^f$ & 0.242$\pm$0.029 & 13.98$^{+0.04}_{-0.05}$ & ...\\
2062.664 & Zn \scriptsize{\uppercase\expandafter{\romannumeral2}} & 0.2560 &2062.556$^g$ & 0.224$\pm$0.031 & 13.47$^{+0.05}_{-0.05}$ & 13.21$^{+0.03}_{-0.03}$\\
2066.1610 & Cr \scriptsize{\uppercase\expandafter{\romannumeral2}} & 0.0515 & 2066.282 & 0.053$\pm$0.022 & 13.46$^{+0.13}_{-0.18}$ & ...\\
2249.8768 & Fe \scriptsize{\uppercase\expandafter{\romannumeral2}} & 0.0018 & 2249.905 & 0.106$\pm$0.025 & 15.17$^{+0.07}_{-0.09}$ & 15.20$^{+0.04}_{-0.04}$\\
2260.7805 & Fe \scriptsize{\uppercase\expandafter{\romannumeral2}} & 0.0024 & 2260.781 & 0.154$\pm$0.022 & 15.22$^{+0.04}_{-0.05}$ & 15.20$^{+0.04}_{-0.04}$\\
2344.2140 & Fe \scriptsize{\uppercase\expandafter{\romannumeral2}} & 0.1140 & 2344.265 & 1.003$\pm$0.043 & 14.62$^{+0.01}_{-0.01}$ & 14.88$^{+0.03}_{-0.03}$\\
2374.4612 & Fe \scriptsize{\uppercase\expandafter{\romannumeral2}} & 0.0313 & 2374.482 & 0.731$\pm$0.031 & 15.00$^{+0.01}_{-0.01}$ & 14.88$^{+0.03}_{-0.03}$\\
2382.7650 & Fe \scriptsize{\uppercase\expandafter{\romannumeral2}} & 0.3200 & 2382.779 & 1.295$\pm$0.034 & 14.33$^{+0.01}_{-0.01}$ & 14.88$^{+0.03}_{-0.03}$\\
2576.8770 & Mn \scriptsize{\uppercase\expandafter{\romannumeral2}} & 0.3508 & 2576.927 & 0.240$\pm$0.023 & 13.14$^{+0.03}_{-0.04}$ & ...\\
2586.6500 & Fe \scriptsize{\uppercase\expandafter{\romannumeral2}} & 0.0691 & 2586.659 & 0.977$\pm$0.034 & 14.73$^{+0.01}_{-0.01}$ & ...\\
2594.4990 & Mn \scriptsize{\uppercase\expandafter{\romannumeral2}} & 0.2710 & 2594.506 & 0.176$\pm$0.022 & 13.09$^{+0.04}_{-0.05}$ & ...\\
2600.1729 & Fe \scriptsize{\uppercase\expandafter{\romannumeral2}} & 0.2390 & 2600.233 & 1.328$\pm$0.031 & 14.42$^{+0.01}_{-0.01}$ & ...\\
2606.4620 & Mn \scriptsize{\uppercase\expandafter{\romannumeral2}} & 0.1927 & 2606.470 & 0.168$\pm$0.020 & 13.20$^{+0.04}_{-0.04}$ & ...\\
2796.3520 & Mg \scriptsize{\uppercase\expandafter{\romannumeral2}} & 0.6123 & 2796.563 & 1.986$\pm$0.022 & 14.22$^{+0.01}_{-0.01}$ & ...\\
2803.5310 & Mg \scriptsize{\uppercase\expandafter{\romannumeral2}} & 0.3054 & 2803.964 & 1.896$\pm$0.022 & 14.49$^{+0.01}_{-0.01}$ & ...\\
2852.9642 & Mg \scriptsize{\uppercase\expandafter{\romannumeral1}} & 1.8100 & 2853.018 & 0.534$\pm$0.022 & 12.81$^{+0.01}_{-0.01}$ & ...\\
\enddata
\tablecomments{Equivalent width is measured in the absorber rest frame at the redshift of 1.5501. Vacuum wavelengths and oscillator strength $f$ are adopted from the Atomic Data conducted by J. X. Prochaska (http://kingpin.ucsd.edu/$\sim$hiresdla/atomic.dat). All statistical uncertainties reported are 1$\sigma$ confidence. However, the systematic error of column densities can exceed 0.05 dex due to continuum fitting and line saturation with ESI data.
}
\tablenotetext{a}{Central wavelengths and equivalent widths are reported in the absorber rest frame.}
\tablenotetext{b}{Column densities are measured by the Apparent Optical Depth (AOD).}
\tablenotetext{c}{Column densities are measured by multi-Voigt fitting.}
\tablenotetext{d}{We assume blend lines at 2026-{\AA} were all Zn~II 2026 when integrating AOD.}
\tablenotetext{e}{We assume blend lines at 2026-{\AA} were all Mg~I 2026 when integrating AOD.}
\tablenotetext{f}{We assume blend lines at 2062-{\AA} were all Cr~II 2062 when integrating AOD.}
\tablenotetext{g}{We assume blend lines at 2062-{\AA} were all Zn~II 2062 when integrating AOD.}
\end{deluxetable}

\clearpage

\begin{deluxetable}{llcccc}
\tabletypesize{\scriptsize}
\tablecaption{Intervening Strong Absorption Lines and Column Density measured by Apparent Optical Depth\label{tbl-2}}
\tablewidth{0pt}
\tablehead{
\colhead{$\lambda_{vacuum}$} & \colhead{Ion} & \colhead{Redshift} & \colhead{EW}
 & \colhead{N$_X$} \\
\colhead{(\AA)} & \colhead{} & \colhead{} & \colhead{(\AA)} & \colhead{log(cm$^{-2}$)
}
}
\startdata
2344.2140 & Fe \scriptsize{\uppercase\expandafter{\romannumeral2}} & 0.7436 & 0.652$\pm$0.200 & 14.48$^{+0.06}_{-0.06}$ \\
2382.7650 & Fe \scriptsize{\uppercase\expandafter{\romannumeral2}} &  & 0.829$\pm$0.210 & 14.12$^{+0.06}_{-0.06}$ \\
2600.1729 & Fe \scriptsize{\uppercase\expandafter{\romannumeral2}} &  & 0.920$\pm$0.211 & 14.32$^{+0.08}_{-0.08}$ \\
2796.3520 & Mg \scriptsize{\uppercase\expandafter{\romannumeral2}} &  & 1.622$\pm$0.186 & 14.06$^{+0.04}_{-0.04}$ \\
2803.5310 & Mg \scriptsize{\uppercase\expandafter{\romannumeral2}} &  & 1.408$\pm$0.207 & 14.03$^{+0.04}_{-0.04}$ \\
\tableline
 & & & & \\
2600.1729 & Fe \scriptsize{\uppercase\expandafter{\romannumeral2}} & 0.5436 & 0.879$\pm$0.418 & 14.13$^{+0.21}_{-0.21}$ \\
2796.3520 & Mg \scriptsize{\uppercase\expandafter{\romannumeral2}} & & 0.923$\pm$0.182 & 13.81$^{+0.30}_{-0.30}$ \\
2803.5310 & Mg \scriptsize{\uppercase\expandafter{\romannumeral2}} & & 0.976$\pm$0.155 & 14.15$^{+0.20}_{-0.20}$ \\
\enddata
\tablecomments{Equivalent width is measured in the rest frame of interested absorber. See note of Table 1 for the oscillator strengths and systematic errors. 
}
\end{deluxetable}

\begin{deluxetable}{lccccc}
%\rotate
\tabletypesize{\normalsize}
\tablecaption{Relative Abundances and Dust Depletions in Associated Absorber\label{tbl-3}}
\tablewidth{0pt}
\tablehead{
\colhead{Element} & \colhead{N$_{adopted}$} & \colhead{[X/Zn]} & \colhead{[X/Zn]$_{SMC}^a$} & \colhead{[X/Zn]$_{MKW}^a$} & \colhead{[X/Zn]$_{MKC}^a$} \\
\colhead{} & \colhead{log(cm$^{-2}$)} & \colhead{} & \colhead{} & \colhead{} & \colhead{}
}
\startdata
Mg..... & $>$14.36 & $>$-1.78 & ... & -0.4 & -0.8 \\
Al$^b$..... & $>$13.83 & $>$-1.15 & ... & -0.9 & -2.0 \\
Si........ & 15.84$^{+0.06}_{-0.06}$ & -0.28$^{+0.07}_{-0.07}$ & 0.07 & -0.3 & -0.9 \\
Cr....... & 13.56$^{+0.06}_{-0.06}$ & -0.69$^{+0.11}_{-0.11}$ & -0.50 & -1.0 & -1.7 \\
Mn...... & 13.15$^{+0.05}_{-0.06}$ & -0.85$^{+0.06}_{-0.06}$ & -0.46 & -0.8 & -0.9 \\
Fe....... & 15.20$^{+0.06}_{-0.06}$ & -0.86$^{+0.08}_{-0.08}$ & -0.57 & -1.2 & -1.8 \\
Ni....... & 14.15$^{+0.08}_{-0.08}$ & -0.69$^{+0.09}_{-0.09}$ & -1.09 & -1.5 & -2.1 \\
Zn....... & 13.21$^{+0.06}_{-0.06}$ & 0.00 & 0.00 & 0.0 & 0.0 \\
\enddata
\tablecomments{The solar photospheric values are adopted from Asplund et al. (2005). For the species with several absorption lines detected, the weighted mean column density of unsaturated lines is used to calculate depletion. The column density measured by multi-Voigt fitting is used when it is available. Errors in column densities are combined quadratically while calculating [X/Zn].}
\tablenotetext{a}{Dust depletion in ``warm" and ``cold" Galactic ISM (MKW and MKC) and SMC ISM were
adopted from Jenkins et al. 1986, Welty et al. 1999 and Welty et al. 2001}
\tablenotetext{b}{No ionization correction was applied, we just use the column density of Al$^+$ here.}
\end{deluxetable}

\begin{deluxetable}{lcccccccccc}
\tabletypesize{\scriptsize}
\tablecaption{Parameters of Optical/UV Extinction Curves\label{tbl-4}}
\tablewidth{0pt}
\tablehead{\colhead{} & \colhead{} & \colhead{$c_1$} & \colhead{$c_2$} & \colhead{$c_3$} & \colhead{$
x_0$} & \colhead{$\gamma$} & \colhead{A$_{bump}$} & \colhead{} & \colhead{} \\
\colhead{Reddened Object} & \colhead{z$_{abs}^a$} & \colhead{(mag)} & \colhead{(mag)} & \colhead{(mag)} & \colhead{($\mu$m$^{-1}$)} & \colhead{($\mu$m$^{-1}$)} & \colhead{} & \colhead{$\chi_{\nu}^2$} & \colhead{Reference}}
\startdata
J003545.13+011441.2 & 1.5501 & -2.17$\pm$0.01 & 0.28$\pm$0.01 & 0.08$\pm$0.01 & 4.59$^b$ & 0.89$^b$ & 0.15$\pm$0.02 & 1.21 & 1\\
J012147.73+002718.7 & 1.3947 & -0.65$\pm$0.02 & 0.06$\pm$0.01 & 0.48$\pm$0.04 & 4.64$\pm$0.01 & 0.80$\pm$0.04 & 0.93$\pm$0.03 & 1.08 & 2,3\\
J085042.21+515911.7 & 1.3265 & -2.70$\pm$0.02 & 0.41$\pm$0.01 & 0.61$\pm$0.07 & 4.54$\pm$0.01 & 1.21$\pm$0.06 & 0.79$\pm$0.05 & 1.14 & 4 \\
J085244.74+343540.4 & 1.3095 & -2.98$\pm$0.02 & 0.47$\pm$0.01 & 0.47$\pm$0.05 & 4.55$\pm$0.01 & 0.84$\pm$0.05 & 0.88$\pm$0.04 & 1.40 & 4\\
J100713.68+285348.4 & 0.8839 & -3.85$\pm$0.05 & 0.65$\pm$0.03 & 6.45$\pm$2.38 & 4.91$\pm$0.15 & 1.78$\pm$0.19 & 5.69$\pm$1.49 & 1.10 & 5\\
J145907.19+002401.2 & 1.3888 & -2.17$\pm$0.02 & 0.08$\pm$0.01 & 9.22$\pm$0.71 & 4.56$\pm$0.02 & 2.68$\pm$0.07 & 5.40$\pm$0.27 & 2.19 & 2,3\\
J160457.50+220300.5 & 1.6405 & -2.09$\pm$0.01 & 0.28$\pm$0.01 & 0.46$\pm$0.03 & 4.58$\pm$0.01 & 0.93$\pm$0.03 & 0.75$\pm$0.03 & 1.37 & 6\\
\enddata
\tablecomments{Best fitted parameters of Optical/UV extinction curves for SDSS 2175-{\AA} absorbers.}
\tablenotetext{a}{Redshift of interested absorber.}
\tablenotetext{b}{These values are fixed during spectrum fitting.}
\tablerefs{(1) this work; (2) Wang et al. 2004; (3) Jiang et al. 2010; (4) Srianand et al. 2008; (5) Zhou et al. 2010; (6) Noterdaeme et al. 2009}
\end{deluxetable}


\begin{references}
\reference{} Abazajian, K. N., \& for the SDSS Collaboration 2009, \apjs,182, 543
\reference{} Antonucci, R. 1993, ARA\&A, 31, 473
\reference{} Asplund, M., Grevesse, N., and Sauval, A. J. 2005, ASPC, 336, 25
\reference{} Bouchet, P., Lequeux, J., Maurice, E., Prevot, L., and Prevot-Burnichon, M. L. 1985, A\&A, 149, 330
\reference{} Brandt, W. N., et al. 2000, \aj, 119, 2349
\reference{} Burbidge, E. M., Lynds, C. R., and Burbidge, G. R. 1966, \apj, 144, 447
\reference{} Cohen, R. D., Burbidge, E. M., Junkkarinen, V. T., Lyons, R. W., and Madejski, G. 1999, BAAS, 31, 942
\reference{} Cui, J., Bechtold, J., Ge, J., \& Meyer, D.M. 2005, ApJ, 633, 649
\reference{} Dong, X. B., Zhou, H. Y., Wang, T. G., Wang, J. X., Li, C. and Zhou, Y. Y. 2005, \apj, 620, 629
\reference{} El{\'i}asd{\'o}ttir, {\'A}., Fynbo, J. P. U., Hjorth, J., Ledoux, C., Watson, D. J., Andersen, A. C., Malesani, D., Vreeswijk, P. M., Prochaska, J. X., Sollerman, J., and Jaunsen, A. O. 2009, \apj, 697, 1725
\reference{} Ellison, S. L., Vreeswijk, P., Ledoux, C., Willis, J. P., Jaunsen, A., Wijers, R. A. M. J., Smette, A., Fynbo, J. P. U., M{\o}ller, P., Hjorth, J., and Kaufer, A. 2006, \mnras, 372, 38
\reference{} Fitzpatrick, E. L. 1989, IAUS, 135, 37
\reference{} Fitzpatrick, E. L., \& Massa, D. 1990, \apjs, 72, 163
\reference{} Fitzpatrick, E. L., \& Massa, D. 2007, \apj, 663, 320
\reference{} Foltz, C. B., Weymann, R. J., Peterson, B, M., Sun, L., Malkan, M. A., and Chaffee, F. H. 1986, \apj, 307, 504
\reference{} Ge, J. Bechtold, J., \& Kulkarni, V.P. 2001, ApJ, 547, L1
\reference{} Ge, J., \& Bechtold, J., 1997, ApJ, 477, L73
\reference{} Gordon, K. D., Clayton, G. C., Misselt, K. A., Landolt, A. U., and Wolff, M. J. 2003, \apj, 594, 279
\reference{} Jenkins, E. B., Savage, B. D., and Spitzer, L. 1986, \apj, 301, 355
\reference{} Jiang, P., Ge, J., Prochaska, J. X., Wang, J., Zhou, H. Y., and Wang, T. G. 2010, \apj, submitted
\reference{} Jiang, P., Ge, J., Zhou, H. Y., Wang, T. G., and Wang, J. X. 2010, in preparation
\reference{} Junkkarinen, V. T., Cohen, R. D., Beaver, E. A., Burbidge, E. M., Lyons, R. W., and Madejski, G. 2004, \apj, 614, 658
\reference{} Li, A. 2007, ASPC, 373, 561
\reference{} M\'{e}nard, B., Nestor, D. B., Turnshek, D. A., Quider, A. M., Richards, G. T., Chelouche, D., and Rao, S. M. 2008, \mnras, 385, 1053
\reference{} Meyer, D. M., and Roth, K. C. 1990, \apj, 363, 57
\reference{} Mushotzky, R. F., Cowie, L. L., Barger, A. J., and Arnaud, K. A. 2000, Nature, 404, 459 
\reference{} Nestor, D. B., Rao, S. M., Turnshek, D. A., and vanden Berk, D. 2003, \apj, 595, 5
\reference{} Noterdaeme, P., Ledoux, C., Petitjean, P., \& Srianand, R. 2008, A\&A, 481, 327
\reference{} Noterdaeme, P., Ledoux, C., Srianand, R., Petitjean, P., and Lopez, S. 2009, A\&A, 503, 765
\reference{} Pei, Y. C. 1992, \apj, 395, 130
\reference{} Pettini, M., Smith, L. J., Hunstead, R. W., and King, D. L. 1994, \apj, 426, 79
\reference{} Pettini, M., Smith, L. J., King, D. L., and Hunstead, R. W. 1997, \apj, 486, 665
\reference{} Pettini, M., Ellison, S. L., Steidel, C. C., and Bowen, D. V. 1999, \apj, 510, 576
\reference{} Pitman, K. M., Clayton, G. C., and Gordon, K. D. 2000, PASP, 112, 537
\reference{} Prochaska, J. X., Sheffer, Y., Perley, D. A., Bloom, J. S., Lopez, L. A., Dessauges-Zavadsky, M., Chen, H.-W., Filippenko, A. V., Ganeshalingam, M., Li, W., Miller, A. A., and Starr, D. 2009, \apj, 691, 27
\reference{} Rao, S. M., Turnshek, D. A., and Nestor, D. B. 2006, \apj, 636, 610
\reference{} Richards, G. T., York, D. G., Yanny, B., Kollgaard, R. I., Laurent-Muehleisen, S. A., and vanden Berk, D. E. 1999, \apj, 513, 576
\reference{} Richards, G. T. et al. 2001, \aj, 121, 2308
\reference{} Richards, G. T. et al. 2003, \aj, 126, 1131
\reference{} Savage, B. D., \& Mathis, J. S. 1979, ARA\&A, 17, 73
\reference{} Savage, B. D., \& Sembach, K. R. 1991, \apj, 379, 245
\reference{} Schlegel, D. J., Finkbeiner, D. P., and Davis, M. 1998, \apj, 500, 525
\reference{} Schmidt, M. 1966, \apj, 144, 443
\reference{} Shanks, T., Georgantopoulos, I., Stewart, G. C., Pounds, K. A., Boyle, B. J., and Griffiths, R. E. 1991, Nature, 353, 315
\reference{} Sheinis, A. I., Bolte, M., Epps, H. W., Kibrick, R. I., Miller, J. S., Radovan, M. V., Bigelow, B. C., and Sutin, B. M. 2002, PASP, 114, 851
\reference{} Srianand, R., Gupta, N., Petitjean, P., Noterdaeme, P., and Saikia, D. J. 2008, \mnras, 391, 69
\reference{} Stoughton, C. et al. 2002, SPIE, 4836, 339 
\reference{} Vanden Berk, D. E. et al. 2001, \aj, 122, 549
\reference{} Wang, J., Hall, P. B., Ge, J., Li, A., and Schenider, D. P. 2004, \apj, 609, 589
\reference{} Welty, D. E., Hobbs, L. M., Lauroesch, J. T., Morton, D. C., Spitzer, L., and York, D. G. 1999, ApJS, 124, 465
\reference{} Welty, D. E., Lauroesch, J. T., Blades, J. C., Hobbs, L. M., and  York, D. G. 2001, \apj, 554, 75
\reference{} Weymann, R. J., Williams, R. E., Peterson, B. M., and Turnshek, D. A. 1979, \apj, 234, 33
\reference{} York, D. G. et al. 2006, \mnras, 367, 945
\reference{} Zhou, H. Y., Ge, J., Lu, H. L., Wang, T. G., Yuan, W. M., Jiang, P., and Shan, H. G. 2010, \apj, 708, 742
\end{references}
\end{document}